\documentclass[a4paper,12pt]{article}
\usepackage[active]{srcltx}
\usepackage{amsmath}\usepackage{amssymb}\usepackage{amscd}

\usepackage{pstricks}

\newcommand{\tsigma}{\tilde{\sigma}}

\newcounter{pic}

\newtheorem{defi}{Definition}\newtheorem{theo}[defi]{Theorem}\newtheorem{prop}[defi]{Proposition}      
\newtheorem{coro}[defi]{Corollary}

\usepackage{empheq}

\begin{document}

$\ $

\vspace{-1.5cm}

\begin{center}
\hfill {\small{CPT-P020-2010}}

\bigskip
{\Large\bf On representations of cyclotomic Hecke algebras}

\vspace{.6cm}

{\large {\bf O. V. Ogievetsky$^{\circ\diamond}$\footnote{On leave of absence from P. N. Lebedev Physical Institute, Leninsky Pr. 53,
117924 Moscow, Russia} and L. Poulain d'Andecy$^{\circ}$}}

\vskip 0.2 cm

$\circ\ ${\large Center of Theoretical Physics\footnote{Unit\'e Mixte de Recherche (UMR 6207) du CNRS et des Universit\'es
Aix--Marseille I, Aix--Marseille II et du Sud Toulon -- Var; laboratoire
affili\'e \`a la FRUMAM (FR 2291)}, Luminy \\
13288 Marseille, France}

\vspace{.5cm}
$\diamond\ ${\large 
J.-V. Poncelet French-Russian Laboratory, UMI 2615 du CNRS, Independent University of Moscow, 11 B. Vlasievski per., 119002 Moscow, Russia
}

\end{center}

\vskip 1cm
\begin{abstract}
An approach, based on Jucys--Murphy elements, to the representation theory of cyclotomic Hecke algebras is developed. The maximality (in the 
cyclotomic Hecke algebra) of the set of the Jucys--Murphy elements is established. A basis of the cyclotomic Hecke algebra is suggested; this basis is 
used to establish the flatness of the deformation without use of the representation theory.
\end{abstract}

\section{{\hspace{-0.6cm}.\hspace{0.55cm}}Introduction}

In the paper \cite{IO2} a Hamiltonian for an open chain model was suggested. This Hamiltonian is written on the universal level, as an element of 
(a certain localization of) the group ring of an "affine Hecke algebra", see \cite{IO2} for details; the model induces an integrable system for any 
representation of the affine Hecke algebra. So, for the study of the model it is important to understand different aspects of the representation theory
of the affine Hecke algebra. In the present article we discuss finite-dimensional representations of the affine Hecke algebra. More precisely, a finite 
dimensional representation of the affine Hecke algebra factors through a certain finite-dimensional quotient of the affine Hecke algebra, a "cyclotomic 
Hecke algebra"; we discuss representations of the cyclotomic Hecke algebras. 
The cyclotomic Hecke algebra is a particular example of a Hecke-type deformation which group rings of Coxeter groups and, more generally, of complex reflection groups admit (the usual Hecke algebra is a deformation of the Coxeter group of type A, that is, the symmetric group). The cyclotomic Hecke 
algebras $H(m,1,n)$ are the Hecke algebras associated to the complex reflection groups of type $G(m,1,n)$; for $m = 1$ and $m = 2$, these are the Hecke 
algebras associated to the Coxeter groups of type A and type B. The cyclotomic Hecke algebras have been introduced in \cite{AK} wherein the representation theory was built and used to prove that the deformation is flat.

Section \ref{sec-def} contains necessary definitions. In Section \ref{sec-norm} we exhibit a basis of the algebra $H(m,1,n)$, quite different from the basis 
presented in \cite{AK}; we use this basis to establish the flatness of the deformation without appealing to the representation theory.

In Sections \ref{ygcyclo}--\ref{sec-comp} an inductive approach to the representation theory of the algebras $H(m,1,n)$ is developed. It uses heavily the 
Jucys-Murphy elements which form a maximal commutative set in the algebras $H(m,1,n)$. This approach extends  the approach developed 
in \cite{OV} for the symmetric groups (generalized to the Hecke algebras in \cite{IO} and to the Brauer algebras and their $q$-deformations in \cite{IO3}). Our
aim is not the construction of the representation theory, it has been already constructed in \cite{AK}. We rather rederive the representation theory directly from the properties of the Jucys-Murphy operators; we encode the representation bases and the Young multi-tableaux by sets of common
eigenvalues of the Jucys-Murphy elements. The irreducible representations of $H(m,1,n)$ are constructed with the help of a new associative algebra; the underlying vector space of this algebra is the tensor product of the cyclotomic Hecke algebra with a free associative algebra generated by
standard $m$-tableaux. The proofs of statements are often omitted in the present letter; details of the results and proofs and also the classical limit of the 
approach will be published in \cite{OL}.

Throughout the paper the ground field is the field $\mathbb{C}$ of complex numbers. 

\setcounter{equation}{0}
\section{{\hspace{-0.6cm}.\hspace{0.55cm}}Cyclotomic Hecke algebras and Jucys-Murphy elements}\label{sec-def}

The braid group of type A (or simply the braid group) on $n$ strands is generated by elements $\sigma_1$, $\dots$, $\sigma_{n-1}$ with the relations:
\begin{empheq}[left=\empheqlbrace]{alignat=4}
\label{def1a}&\sigma_i\sigma_{i+1}\sigma_i=\sigma_{i+1}\sigma_i\sigma_{i+1} &\hspace{1cm}& \textrm{for all $i=1,\dots,n-2$\ ,}\\[.5em]
\label{def1b}&\sigma_i\sigma_j=\sigma_j\sigma_i && \textrm{for all $i,j$ such that $|i-j|>1$\ .}
\end{empheq}

The braid group of type B (sometimes called affine braid group)
is obtained by adding to the generators $\sigma_i$ the generator $\tau$ with the relations:
\begin{empheq}[left=\empheqlbrace]{alignat=3}
\label{def1'a}&\tau\sigma_1\tau\sigma_1=\sigma_1\tau\sigma_1\tau\ ,&\\[.5em]
\label{def1'b}&\tau\sigma_i=\sigma_i\tau & \textrm{for $i>1$\ .}
\end{empheq}

The elements $J_i$, $i=1,\dots,n$, of the braid group of type B defined inductively by the following initial condition and recursion:
\begin{equation}\label{JM}J_1=\tau\ ,\quad J_{i+1}=\sigma_iJ_i\sigma_i\ ,\end{equation}
are called Jucys-Murphy elements.
It is well known that they form a commutative set of elements. In addition,  $J_i$ commutes with all $\sigma_k$ except $\sigma_{i-1}$ and 
$\sigma_i$,
\begin{equation}\label{JMcomm}J_i\sigma_k=\sigma_k J_i\ \ \ {\textrm{if}}\ \  k>i\ \  {\textrm{and if}}\ \  k<i-1\ .\end{equation}

The affine Hecke algebra $\hat{H}_n$ is the quotient of the group algebra of the braid group of type B by:
\begin{equation}\label{def1''b}
\sigma_i^2=(q-q^{-1})\sigma_i+1\qquad  \textrm{for all $i=1,\dots,n-1$\ .}
\end{equation}
The usual Hecke algebra $H_n$ is the algebra generated by elements $\sigma_1$, $\dots$, $\sigma_{n-1}$ with the relations
(\ref{def1a})--(\ref{def1b}) and (\ref{def1''b}).

The cyclotomic Hecke algebra $H(m,1,n)$ is the quotient of the affine Hecke algebra $\hat{H}_n$ by
\begin{equation}\label{def1''a}
(\tau-v_1)\dots(\tau-v_m)=0\ .\end{equation}
The algebra $H(m,1,n)$ is a deformation of the group algebra $\mathbb{C}G(m,1,n)$ of the complex reflection group $G(m,1,n)$.
In particular, $H(1,1,n)$ is isomorphic to the Hecke algebra of type A and $H(2,1,n)$ is isomorphic to the Hecke algebra of type B. 

\vskip .2cm
The group $G(m,1,n)$ is isomorphic to $S_n\wr C_m$, the wreath product of the symmetric group $S_n$ with the cyclic group of $m$ elements. 

\vskip .2cm
The specialization  of $H(m,1,n)$ is semi-simple if and only if the numerical values of the parameters satisfy (see \cite{Ari-sim})
\begin{equation}\label{sesi1}
1+q^2+\dots+q^{2N}\neq 0\ \ {\textrm{for all}}\ \ N:\ N<n\ \end{equation} 
and
\begin{equation}\label{sesi2}
q^{2i}v_j-v_k\neq 0\ \ {\textrm{for all}}\ \ i,j,k \ \ {\textrm{such that}}\ \  j\neq k \ {\textrm{and}}\ -n<i<n\ .\end{equation}
In the sequel we work either with a generic cyclotomic Hecke algebra (that is, $v_1$, $\dots$, $v_m$ and $q$ are indeterminates) or in the semi-simple situation with an additional requirement:
\begin{equation}\label{sesi3} v_j\neq 0\ ,\ j=1,\dots,m\ .\end{equation}

As $n$ varies, the algebras $H(m,1,n)$ form an ascending chain of algebras :
\begin{equation}\label{chaine}\{e\}\subset H(m,1,0)\subset H(m,1,1)\subset\dots\subset H(m,1,n)\subset\dots.\end{equation}

\vskip .2cm
We shall denote by the same symbols $J_i$ the images of the Jucys-Murphy elements in the cyclotomic Hecke algebra. As a by-product of our
construction, we shall see that the set of  the Jucys-Murphy elements $\{ J_1,\dots ,J_n\}$ is maximal commutative in the generic algebra 
$H(m,1,n)$; more precisely, the algebra of polynomials in Jucys-Murphy elements coincides with the algebra generated by the union of the centers 
of $H(m,1,k)$ for $k=1,\dots,n$.

\setcounter{equation}{0}
\section{{\hspace{-0.6cm}.\hspace{0.55cm}}Normal form for $H(m,1,n)$}\label{sec-norm}

We exhibit a basis for the algebra $H(m,1,n)$. Several known facts about the chain (with respect to $n$) of the algebras $H(m,1,n)$ are reestablished with the help of this basis. In particular, we show that $H(m,1,n)$ is a flat deformation of the group ring of $G(m,1,n)$. This was proved in \cite{AK} with the use of the representation theory.

\vskip .2cm
Let $\tilde{W}$ be the subalgebra of $H(m,1,n)$ generated by the elements $\tau,\sigma_1,\dots,\sigma_{n-2}$.

\begin{prop}
{\hspace{-.2cm}.\hspace{.2cm}}\label{normalform-H} 
If $\tilde{\mathcal{B}}$ is a linear basis of $\tilde{W}$ then the elements
\begin{equation}\label{normalform2}
\sigma_j^{-1}\sigma_{j-1}^{-1}\dots \sigma_1^{-1}\tau^{\alpha}\sigma_1\sigma_2\dots \sigma_{n-1}\tilde{b}\ ,
\end{equation}
where $j\in\{0,\dots,n-1\}$, $\alpha\in\{0,\dots,m-1\}$ and $\tilde{b}$ is an 
element of $\tilde{\mathcal{B}}$, form a linear basis of $H(m,1,n)$.
\end{prop}

\begin{coro}
{\hspace{-.2cm}.\hspace{.2cm}}\label{dimension} 
(i) The algebra $H(m,1,n)$ is a flat deformation of the group ring $\mathbb{C}G(m,1,n)$; in other words, $H(m,1,n)$ is a free 
 $\mathbb{C}[q,q^{-1},v_1,\dots ,v_m]$-module of dimension
\begin{equation}\label{dim-H}
\dim \bigl(H(m,1,n)\bigr)=|G(m,1,n)|=n!m^n\ .
\end{equation}
(ii) Moreover the subalgebra $\tilde{W}$ is isomorphic to $H(m,1,n-1)$.
\end{coro}

\bigskip
We can construct recursively a global normal form for elements of $H(m,1,n)$ using 
the Proposition \ref{normalform-H} and the Corollary \ref{dimension}, statement (ii). Let $R_k$ be the following set of elements: 
\[R_k=\{\sigma_j^{-1}\sigma_{j-1}^{-1}\dots \sigma_1^{-1}\tau^{\alpha}\sigma_1\sigma_2\dots \sigma_{k-1},\ j=0,\dots,k-1,\ \alpha=0,\dots,m-1\}.\]

\begin{coro}
{\hspace{-.2cm}.\hspace{.2cm}}\label{normalform-fin} 
Any element $x\in H(m,1,n)$ can be written uniquely as a linear combination of elements 
$u_nu_{n-1}\dots u_1$ where $u_k\in R_k$ for $k=1,\dots,n$. 
\end{coro}

\setcounter{equation}{0}
\section{{\hspace{-0.6cm}.\hspace{0.55cm}}Spectrum of Jucys--Murphy elements and Young $m$-tableaux}\label{ygcyclo}

We begin to develop an approach, based on the Jucys-Murphy elements, to the representation theory of the chain (with respect to $n$) of the 
cyclotomic Hecke algebras $H(m,1,n)$. This is a generalization of the approach of \cite{OV}.

\paragraph{1.} The first step consists in construction of all representations of $H(m,1,n)$ verifying two conditions. First, the Jucys-Murphy elements  
$J_1,\dots,J_n$ are represented by semi-simple (diagonalizable) operators. Second, for every $i=1,\dots,n-1$ the action of the sub-algebra generated 
by $J_i$, $J_{i+1}$ and $\sigma_i$ is completely reducible. We shall use the name $C$-representations ($C$ is the first letter in "completely reducible") 
for these representations.  At the end of the construction we shall see that all irreducible representations of $H(m,1,n)$ are $C$-representations.

\vskip .2cm
Following \cite{OV} we denote by ${\mathrm{Spec}}(J_1,\dots,J_n)$ the set of strings of eigenvalues of the Jucys-Murphy elements in the set of $C$-representations: $\Lambda=(a^{(\Lambda)}_1,\dots,a^{(\Lambda)}_n)$ belongs to the set ${\mathrm{Spec}}(J_1,\dots,J_n)$ if there is a vector 
$e_{\Lambda}$ in the space of some $C$-representation such that 
$J_i(e_{\Lambda})=a^{(\Lambda)}_ie_{\Lambda}$ for all $i=1,\dots,n$. Every $C$-representation possesses a basis formed by vectors $e_{\Lambda}$
(this is a reformulation of the first condition in the definition of $C$-representations).
Since $\sigma_k$ commutes with $J_i$ for $k>i$ and $k<i-1$, the action of $\sigma_k$ on a vector  $e_{\Lambda}$, 
 $\Lambda\in {\mathrm{Spec}}(J_1,\dots,J_n)$, is "local" in the sense that $\sigma_k(e_{\Lambda})$ is a linear combination of $e_{\Lambda'}$ such that 
$a^{(\Lambda')}_i=a^{(\Lambda)}_i$ for $i\neq k,k+1$.

\paragraph{2. Affine Hecke algebra $\hat{H}_2$.} Consider the affine Hecke algebra $\hat{H}_2$, which is generated by $X$, $Y$ and $\sigma$ with the relations:
\begin{equation}\label{affHec}XY=YX,\quad Y=\sigma X\sigma,\quad \sigma^2=(q-q^{-1})\sigma+1\ .\end{equation}
It is realized by $J_i$, $J_{i+1}$ and $\sigma_i$ for all $i=1,\dots,n-1$. We reproduce here the result of \cite{IO} concerning the classification of irreducible representations with diagonalizable $X$ and $Y$ of $\hat{H}_2$.

\vskip .2cm
There are one-dimensional and two-dimensional irreducible representations.
\begin{itemize}
 \item The one-dimensional irreducible representations are given by
\begin{equation}\label{mat-d1}
X\mapsto a,\quad Y\mapsto q^{\pm2}a,\quad\sigma\mapsto \pm q^{\pm1}\ .
\end{equation}
\item The two-dimensional irreducible representations are given by
\[\sigma\mapsto\left(\begin{array}{cc}0 & 1\\ 1 & q-q^{-1}\end{array}\right),\quad X\mapsto\left(\begin{array}{cc}a & -(q-q^{-1})b\\ 0 & b\end{array}\right),\quad Y\mapsto\left(\begin{array}{cc}b & (q-q^{-1})b\\ 0 & a\end{array}\right),\]
with $b\neq a$ in order for $X$ and $Y$ to be diagonalizable and with $b\neq q^{\pm2}a$ to ensure irreducibility. By a change of basis we transform $X$ and $Y$ to a diagonal form:
\begin{equation}\label{mat-d2}
\sigma\mapsto\left(\begin{array}{cc}\frac{(q-q^{-1})b}{b-a}\ &\ 1-\frac{(q-q^{-1})^2ab}{(b-a)^2}\\ 1 \ &\ -\frac{(q-q^{-1})a}{b-a}\end{array}\right),\quad X\mapsto\left(\begin{array}{cc}a &0\\ 0 & b\end{array}\right),\quad Y\mapsto\left(\begin{array}{cc}b & 0\\ 0 & a\end{array}\right).
\end{equation}
\end{itemize}

\paragraph{3.} We return to strings of eigenvalues of Jucys-Murphy elements.

\begin{prop}
{\hspace{-.2cm}.\hspace{.2cm}}
 \label{prop2}
Let $\Lambda=(a_1,\dots,a_i,a_{i+1},\dots,a_n)\in {\mathrm{Spec}}(J_1,\dots,J_n)$ and let  $e_{\Lambda}$ be a corresponding vector. Then: 
\begin{itemize}
\item[(a)] We have $a_i\neq a_{i+1}$.
\item[(b)] If $a_{i+1}=q^{\pm2}a_i$ then $\sigma_i(e_{\Lambda})=\pm q^{\pm1}e_{\Lambda}$. 
\item[(c)] If $a_{i+1}\neq q^{\pm2}a_i$ then 
$\Lambda'=(a_1,\dots,a_{i+1},a_i,\dots,a_n)\in {\mathrm{Spec}}(J_1,\dots,J_n)$; moreover, the vector 
$\sigma_i(e_{\Lambda})-\frac{(q-q^{-1})a_{i+1}}{a_{i+1}-a_i}e_{\Lambda}$ corresponds to the string $\Lambda'$ (see (\ref{mat-d2}) with $b=a_{i+1}$ and $a=a_i$). 
\end{itemize}
\end{prop}

\paragraph{4. Content strings.}

\begin{defi}
{\hspace{-.2cm}.\hspace{.2cm}}
 \label{def-cont} A content string $(a_1,\dots,a_n)$ is a string of numbers satisfying the following conditions:
\begin{itemize}
\item[(c1)]  $a_1\in\{v_1,\dots ,v_m\}$;
\item[(c2)]  for all $j>1$: if $a_j=v_k q^{2z}$ for some $k$ and $z\neq 0$, then $\{v_k q^{2(z-1)},v_k q^{2(z+1)}\}\cap\{a_1,\dots,a_{j-1}\}$ is non-empty;
\item[(c3)]  if $a_i\!=\!a_j\!=\!v_kq^{2z}$ with $i<j$ for some $k$ and $z$, then $\{v_k q^{2(z-1)},v_k q^{2(z+1)}\}\!\subset\!\{a_{i+1},\dots,a_{j-1}\}$.
\end{itemize}

\vskip .2cm
The set of content strings of length $n$ we denote by ${\mathrm{Cont}}_m(n)$.
\end{defi}

Here is the "cyclotomic" analogue of the Theorem 5.1 in \cite{OV} and the Proposition 4 in \cite{IO}.

\begin{prop}
{\hspace{-.2cm}.\hspace{.2cm}}
 \label{prop3}
If a string of numbers $(a_1,\dots,a_n)$ belongs to ${\mathrm{Spec}}(J_1,\dots,J_n)$ then it belongs to ${\mathrm{Cont}}_m(n)$.
\end{prop}

\paragraph{5. Young $m$-diagrams and $m$-tableaux.} A Young $m$-diagram, or $m$-partition, is an $m$-tuple of Young diagrams $\lambda^{(m)}=(\lambda_1,\dots,\lambda_m)$. The length of a Young diagram $\lambda$ 
is the number of nodes of the diagram and is denoted by $|\lambda|$. By definition the length of an $m$-tuple $\lambda^{(m)}=(\lambda_1,\dots,\lambda_m)$ is 
\begin{equation}\label{lenmdi}|\lambda^{(m)}|:=|\lambda_1|+\dots+|\lambda_m|\ .\end{equation}

Let the length of the $m$-tuple be $n$. We place the numbers $1,\dots,n$ in the nodes of these diagrams in such a way that in every diagram the 
numbers in the nodes are in ascending order along rows and columns in right and down directions. This is a standard Young $m$-tableau of 
shape $\lambda^{(m)}$. 

\vskip .2cm
We associate to each node of a Young $m$-diagram a number (the "content") which is $v_kq^{2(s-r)}$ for the node in the line $r$ and column $s$ in the $k^{th}$ tableau of the $m$-tuple. To a standard Young $m$-tableau we associate a string of contents: the position number $i$ of a string contains the content
of the node, labeled by $i$, of the standard Young $m$-tableau. 
 
\begin{prop}
{\hspace{-.2cm}.\hspace{.2cm}}
 \label{prop4}
This association establishes a bijection between the set of standard Young $m$-tableaux of length $n$ and the set ${\mathrm{Cont}}_m(n)$. 
\end{prop}

\setcounter{equation}{0}
\section{{\hspace{-0.6cm}.\hspace{0.55cm}}Construction of representations}\label{ygcyclo'}

We proceed as in \cite{OP}. We first define an algebra structure on a tensor product of the algebra $H(m,1,n)$ with a free associative algebra
generated by the standard $m$-tableaux corresponding to $m$-partitions of $n$. Then, by 
evaluation (with the help of the simplest one-dimensional representation of $H(m,1,n)$) from the right, we build representations. 

\bigskip
Define, for any $\sigma_i$ among the generators $\sigma_1,\dots,\sigma_{n-1}$ of $H(m,1,n)$, the Baxterized elements $\sigma_i(\alpha,\beta)$ by 
\begin{equation}\label{bax-sig}
\sigma_i(\alpha,\beta):=\sigma_i+(q-q^{-1})\frac{\beta}{\alpha-\beta}\ .
\end{equation} 
The parameters $\alpha$ and $\beta$ are called spectral parameters. We recall some useful relations for the Baxterized generators $\sigma_i$.

\begin{prop}
 {\hspace{-.2cm}.\hspace{.2cm}}
 \label{prop-bax}
The following relations hold:
\begin{equation}\label{rel-bax}
\begin{array}{c}\sigma_i(\alpha,\beta)\sigma_i(\beta,\alpha)=f(\alpha,\beta)f(\beta,\alpha)\qquad\quad\textrm{with\  $f(\alpha,\beta)=\frac{q\alpha-q^{-1}\beta}{\alpha-\beta}$,}\\[0.7em]
\sigma_i(\alpha,\beta)\sigma_{i+1}(\alpha,\gamma)\sigma_i(\beta,\gamma)=\sigma_{i+1}(\beta,\gamma)\sigma_i(\alpha,\gamma)\sigma_{i+1}(\alpha,\beta),\\[0.85em]
\sigma_i(\alpha,\beta)\sigma_j(\gamma,\delta)=\sigma_j(\gamma,\delta)\sigma_i(\alpha,\beta)\qquad\quad\textrm{if\ \  $|i-j|>1$.}
\end{array}
\end{equation}
\end{prop}

\bigskip
Let $\lambda^{(m)}$ be an $m$-partition of length $n$. Consider a set of free generators labeled by standard $m$-tableaux of shape 
$\lambda^{(m)}$; for a standard $m$-tableau $X_{\lambda^{(m)}}$ we denote by $\mathcal{X}_{\lambda^{(m)}}$ the corresponding free generator 
and by $c(X_{\lambda^{(m)}}|i)$ the content  (see the preceding Section for the definition) of the node carrying the number $i$.
 
\begin{prop}
{\hspace{-.2cm}.\hspace{.2cm}}
 \label{prop-rel}
The relations
\begin{equation}\label{rel-a1}
\Bigl(\sigma_i+\frac{(q-q^{-1})
c(X_{\lambda^{(m)}}|i+1)}{c(X_{\lambda^{(m)}}|i)-c(X_{\lambda^{(m)}}|i+1)}\Bigr)\cdot\mathcal{X}_{\lambda^{(m)}}=\mathcal{X}_{\lambda^{(m)}}s_i\cdot\Bigl(\sigma_i+\frac{(q-q^{-1})c(X_{\lambda^{(m)}}|i)}{c(X_{\lambda^{(m)}}|i+1)-c(X_{\lambda^{(m)}}|i)}\Bigr)\end{equation}
and
\begin{equation}\label{rel-a2}
\Bigl(\tau-c(X_{\lambda^{(m)}}|1)\Bigr)\mathcal{X}_{\lambda^{(m)}}=0
\end{equation}
are compatible with the relations for the generators $\tau,\sigma_1,\dots,\sigma_{n-1}$ of the algebra $H(m,1,n)$. The element 
$\mathcal{X}_{\lambda^{(m)}}s_i$ corresponds to the $m$-tableau $X_{\lambda^{(m)}}s_i$ obtained from $X_{\lambda^{(m)}}$ by exchanging the nodes with numbers 
$i$ and $(i+1)$. The tableau $X_{\lambda^{(m)}}s_i$ is of the same (as $X_{\lambda^{(m)}}$) shape $\lambda^{(m)}$. 
If the resulting $m$-tableau $X_{\lambda^{(m)}}s_i$ is not standard, we put $\mathcal{X}_{\lambda^{(m)}}s_i=0$. 
\end{prop}

We explain the meaning of the word "compatible" used in the formulation of the Proposition.

\vskip .2cm
Let ${\cal{F}}$ be the free associative algebra
generated by $\tilde{\tau}$, $\tsigma_1,\dots,\tsigma_{n-1}$.  The algebra $H(m,1,n)$ is naturally the quotient of ${\cal{F}}$. 

\vskip .2cm
Let $\mathbb{C}[\mathcal{X}]$ be 
the free associative algebra whose generators  
$\mathcal{X}_{\lambda^{(m)}}$  range over all standard $m$-tableaux of shape $\lambda^{(m)}$ for all $m$-partitions $\lambda^{(m)}$ of $n$.

\vskip .2cm
Consider an algebra structure on the space $\mathbb{C}[\mathcal{X}]\otimes {\cal{F}}$ for which: (i) the map $\iota_1:x\mapsto x\otimes 1$, 
$x\in \mathbb{C}[\mathcal{X}]$, is an isomorphism of $\mathbb{C}[\mathcal{X}]$ with its image with respect to $\iota_1$; (ii) the map 
$\iota_2:\phi\mapsto 1\otimes \phi$, $\phi\in {\cal{F}}$, is an isomorphism of ${\cal{F}}$ with its image with respect to $\iota_2$; (iii) 
the formulas (\ref{rel-a1})-(\ref{rel-a2}), extended by associativity, provide the rules to rewrite elements of the form $(1\otimes \phi)(x\otimes 1)$, $x\in \mathbb{C}[\mathcal{X}]$,
$\phi\in {\cal{F}}$, as elements of $\mathbb{C}[\mathcal{X}]\otimes {\cal{F}}$. 

\vskip .2cm
The "compatibility" means that we have an induced structure of an associative algebra on the space $\mathbb{C}[\mathcal{X}]\otimes H(m,1,n)$. 
More precisely, if we multiply any relation of the cyclotomic Hecke algebra $H(m,1,n)$ (the relation is viewed as an element of the free algebra 
${\cal{F}}$) from the right by a generator $\mathcal{X}_{\lambda^{(m)}}$ (this is a combination of the form "relation of $H(m,1,n)$ times 
$\mathcal{X}_{\lambda^{(m)}}$") and use the "instructions" (\ref{rel-a1})-(\ref{rel-a2}) to move all appearing $\mathcal{X}$'s to the left (the free generator 
changes but the expression stays always linear in $\mathcal{X}$) then we obtain a combination of terms of the form "$\mathcal{X}$ times 
relation of $H(m,1,n)$".

\bigskip
Let $\vert\rangle$ be a ``vacuum" - a basic vector of a one-dimensional $H(m,1,n)$-module; for example, $\sigma_i\vert\rangle=q\vert\rangle$ for all $i$ and
$\tau\vert\rangle =v_1\vert\rangle$. Moving, in the expressions $\phi\mathcal{X}_{\lambda^{(m)}}\vert\rangle$, $\phi\in H(m,1,n)$, the elements 
$\mathcal{X}$'s to the left and using the module structure, we build, due to the compatibility, a representation of $H(m,1,n)$ on the vector space 
$V_{\lambda^{(m)}}$ spanned by
$\mathcal{X}_{\lambda^{(m)}}\vert\rangle$. We shall, by a slight abuse of notation, denote the symbol $\mathcal{X}_{\lambda^{(m)}}\vert\rangle$ again
by $\mathcal{X}_{\lambda^{(m)}}$. This procedure leads to the following formulas for the action of the generators $\tau,\sigma_1,\dots,\sigma_{n-1}$ on 
the basis vectors $\{\mathcal{X}_{\lambda^{(m)}}\}$ of $V_{\lambda^{(m)}}$:
\begin{equation}\label{rep-a1}\begin{array}{rcl}
\sigma_i\,:\,\mathcal{X}_{\lambda^{(m)}}&\mapsto& -\ (q-q^{-1}){\displaystyle \frac{c(X_{\lambda^{(m)}}|i+1)}{c(X_{\lambda^{(m)}}|i)-c(X_{\lambda^{(m)}}|i+1)}}\mathcal{X}_{\lambda^{(m)}}\\[2em]
&&+\ {\displaystyle \frac{qc(X_{\lambda^{(m)}}|i+1)-q^{-1}c(X_{\lambda^{(m)}}|i)}{c(X_{\lambda^{(m)}}|i+1)-
c(X_{\lambda^{(m)}}|i)}}\mathcal{X}_{\lambda^{(m)}}s_i\end{array}\end{equation}
and
\begin{equation}\label{rep-a2}
\tau\,:\,\mathcal{X}_{\lambda^{(m)}}\mapsto c(X_{\lambda^{(m)}}|1)\mathcal{X}_{\lambda^{(m)}}\ .
\end{equation}
As before, it is assumed here that $\mathcal{X}_{\lambda^{(m)}}s_i=0$ if $X_{\lambda^{(m)}}s_i$ is not a standard $m$-tableau.

\setcounter{equation}{0}
\section{{\hspace{-0.6cm}.\hspace{0.55cm}}Completeness}\label{sec-comp}

We conclude the study of representations of $H(m,1,n)$. First, with the construction of the preceding Section, we complete the description of the set ${\mathrm{Spec}}(J_1,\dots,J_n)$ started in Section \ref{ygcyclo}. Then we check (by calculating the sum of squares of dimensions) 
that we obtain within this approach all irreducible representations of the algebra $H(m,1,n)$.

\vskip .2cm
We underline that the restrictions (\ref{sesi1})--(\ref{sesi3}) are essential for the statements below. 

\begin{prop}
{\hspace{-.2cm}.\hspace{.2cm}}
 \label{prop-cont-spec}
The set ${\mathrm{Spec}}(J_1,\dots,J_n)$, the set ${\mathrm{Cont}}_m(n)$ and the set of standard $m$-tableaux are in pairwise bijections.
\end{prop}

\begin{theo}
{\hspace{-.2cm}.\hspace{.2cm}}
 \label{prop-fin}
The representations $V_{\lambda^{(m)}}$ (labeled by the $m$-partitions) of the algebra $H(m,1,n)$ constructed in the Section \ref{ygcyclo'} are irreducible and pairwise non-isomorphic. Moreover the sum of the squares of the dimensions of the constructed representations equals the dimension of $H(m,1,n)$.
\end{theo}

In particular, the Jucys--Murphy elements distinguish basis elements in the sum of all irreducible representations; therefore the commutative set of 
Jucys--Murphy elements is maximal in the algebra $H(m,1,n)$. 

\vskip .2cm
As a by-product we obtain that the algebra $H(m,1,n)$ is semi-simple under the restrictions (\ref{sesi1})--(\ref{sesi3}).


\begin{thebibliography}{99}\addcontentsline{toc}{section}{References}

\bibitem{AK} Ariki S. and Koike K., \emph{A Hecke algebra of $(\mathbb{Z}/r\mathbb{Z})\wr S_n$ and construction of its irreducible representations}, Adv. 
in Math. \textbf{106} (1994) 216--243.

\bibitem{Ari-sim} Ariki S., \emph{On the semi-simplicity of the Hecke algebra of $(\mathbb{Z}/r\mathbb{Z})\wr S_n$}, J. Algebra \textbf{169} (1994) 216--225.

\bibitem{IO2} Isaev A. P. and Ogievetsky O. V., \emph{On Baxterized solutions of reflection equation and integrable chain models}, Nucl. Phys. 
B \textbf{760} (2007) 167--183. ArXiv: math-ph/0510078.

\bibitem{IO} Isaev A. P. and Ogievetsky O. V., \emph{On representations of Hecke algebras}, Czech. Journ. Phys. \textbf{55} No. 11 (2005) 1433--1441.

\bibitem{IO3} Isaev A. P. and Ogievetsky O. V., \emph{Jucys-Murphy elements for Birman-Murakami-Wenzl algebras}, Proc. of Int. 
Workshop "Supersymmetries and Quantum Symmetries", Dubna (2009). ArXiv: 0912.4010 [math.QA] 

\bibitem{OL} Ogievetsky O. V. and Poulain d'Andecy L., \emph{Cyclotomic Hecke algebras: Jucys-Murphy elements, representations, classical limit}, to appear.

\bibitem{OP} Ogievetsky O. and Pyatov P., \emph{Lecture on Hecke algebras}, in Proc. of the Int. School ``Symmetries and Integrable Systems'', Dubna (1999).

\bibitem{OV} Okounkov A. and Vershik A., \emph{A new approach to representation theory of symmetric groups II},
Selecta Math (New series) \textbf{2} No. 4 (1996) 581--605.


\end{thebibliography}
\end{document}